\documentclass{wlscirep}

\usepackage[utf8]{inputenc}
\usepackage[T1]{fontenc}
\usepackage{bm}
\usepackage{multicol}
\usepackage{float}
\usepackage{multirow}
\usepackage[mathlines]{lineno}
\usepackage{adjustbox}
\title{Determining the Proton's Gluonic Gravitational Form Factors}
\author[3,1]{B.~Duran}
\author[1,3**]{Z.-E.~Meziani}
\author[1]{S.~Joosten}
\author[2]{M.~K.~Jones}
\author[1]{S.~Prasad}
\author[1]{C.~Peng}
\author[1]{W.~Armstrong}
\author[3]{H.~Atac}
\author[2]{E.~Chudakov}
\author[5]{H.~Bhatt}
\author[5]{D.~Bhetuwal}
\author[11]{M.~Boer}
\author[2]{A.~Camsonne}
\author[2]{J.-P.~Chen}
\author[2]{M.~M.~Dalton}
\author[3]{N.~Deokar}
\author[2]{M.~Diefenthaler}
\author[5]{J.~Dunne}
\author[5]{L.~El~Fassi}
\author[9]{E.~Fuchey}
\author[4]{H.~Gao}
\author[2]{D.~Gaskell}
\author[2]{O.~Hansen}
\author[6]{F.~Hauenstein}
\author[2]{D.~Higinbotham}
\author[3]{S.~Jia}
\author[5]{A.~Karki}
\author[2]{C.~Keppel}
\author[7]{P.~King}
\author[10]{H.S.~Ko}
\author[4]{X.~Li}
\author[3]{R.~Li}
\author[2]{D.~Mack}
\author[2]{S.~Malace}
\author[2]{M.~McCaughan}
\author[8]{R.~E.~McClellan}
\author[2]{R.~Michaels}
\author[2]{D.~Meekins}
\author[3]{M.~Paolone}
\author[2]{L.~Pentchev}
\author[2]{E.~Pooser}
\author[9]{A.~Puckett}
\author[7]{R.~Radloff}
\author[3]{M.~Rehfuss}
\author[1]{P.~E.~Reimer}
\author[1]{S.~Riordan}
\author[2]{B.~Sawatzky}
\author[4]{A.~Smith}
\author[3]{N.~Sparveris}
\author[2]{H.~Szumila-Vance}
\author[2]{S.~Wood}
\author[1]{J.~Xie}
\author[1]{Z.~Ye}
\author[6]{C.~Yero}
\author[4]{Z.~Zhao}

\affil[List of affiliations]{\footnote{$^1$Argonne National Laboratory, Lemont, IL 60439, USA.$^2$Thomas Jefferson National Accelerator Facility, Newport News, VA 23606, USA.$^3$Temple University, Philadelphia, PA 19122, USA. $^4$Duke University, Durham 27708, NC, USA. $^5$ Mississippi State University, Mississippi State 39762, MS, USA. $^6$Old dominion University, Norfolk 23529, VA, USA. $^7$Ohio University, Athens 45701, OH, USA. $^8$ Pensacola State College, Pensacola, FL 32504, $^9$University of Connecticut, Storrs 06269, CT, USA. $^{10}$Universit\'e Paris-Saclay, Gif-sur-Yvette 91190, Essone France. $^{11}$Virginia Polytechnic Institute \& State University, Blacksburg, VA 24061, USA. $^{**}$corresponding author:zmeziani@anl.gov}}

\begin{abstract}
The proton is one of the main building blocks of all visible matter in the
  universe\cite{NASRep:2018}. Among its intrinsic properties are its electric charge,
  mass, and spin~\cite{Workman:2022ynf}. These emerge from the complex dynamics of its
  fundamental constituents, quarks and gluons, described by the theory of quantum
  chromodynamics (QCD)~\cite{Shifman:1978bx,Shifman:1978by,Shifman:1978zn}. Using electron
  scattering, its electric charge and spin, shared among the quark constituents, have been
  the topic of active investigation\cite{Workman:2022ynf}. An example is the novel precision measurement of the proton’s electric charge radius~\cite{Xiong:2019umf}. In contrast, little is known about the proton’s inner mass density, dominated by the energy carried by the gluons, which are hard to access through electron scattering since gluons carry no electromagnetic charge. Here, we chose to probe this gluonic gravitational density using a small color dipole, the $J/\psi$ particle, through its threshold photoproduction. From our data, we determined, for the first time, the proton’s gluonic gravitational form factors~\cite{Pagels:1966zza,Teryaev:2016edw}. We used a variety of models~\cite{Kharzeev:2021qkd,Guo:2021ibg,Mamo:2019mka} and determined, in all cases, a mass radius that is notably smaller than the electric charge radius. In some cases, the determined radius, although model dependent, is in excellent agreement with first-principle predictions from lattice QCD~\cite{Pefkou:2021fni}. This work paves the way for a deeper understanding of the salient role of gluons in providing gravitational mass to visible matter.  

\end{abstract}

\begin{document}
\flushbottom
\maketitle
\thispagestyle{empty}
\begin{multicols}{2}

In the standard model of cosmology, after the big bang nucleosynthesis, most of the mass of the visible universe was encapsulated in protons, neutrons, and nuclei. In the preceding hadron epoch, the color-neutral confined systems, such as the proton and neutron, resulted from the interplay of strong color forces among the fundamental constituents, quarks, and gluons. Surprisingly, emerging from this interplay is a total nucleon mass much larger than the constituents' mass sum. Thus, the origin of the proton mass is an essential puzzle piece in our understanding of the structure of visible matter in the universe~\cite{NASRep:2018}. 
The triumphant discovery of the Higgs boson offered a crucial explanation for the origin of quark masses. However, the quarks are almost massless (few MeV) and account only for a tiny fraction of the total proton mass of about 1 GeV, even when accounting for their relativistic nature. 
Thus the question arises: How do the massless gluons provide the sizeable remaining mass of the proton, and how is this mass distributed across the confinement size of the proton? 
While Einstein's original definition of the mass of a body, m= E/c$^2$ starts to answer this question, we can gain real insight through the measurement of the proton's gravitational form factors (GFFs) and the determination of the trace anomaly. 
GFFs are the matrix elements of the proton's energy-momentum tensor (EMT)\cite{Pagels:1966zza,Teryaev:2016edw} and encode its mechanical properties, while the trace anomaly of the EMT is a key component of the origin of mass according to Quantum Chromodynamics (QCD)~\cite{Shifman:1978bx,Shifman:1978by,Shifman:1978zn}.
Moreover, with the advent of lattice QCD, we can challenge and benchmark our understanding of the proton's internal structure with ab-initio calculations~\cite{Wilson:1974sk,Durr:2008zz,Borsanyi:2014jba}.

\begin{figure*}[b!]
     \centering
    \includegraphics[width=17cm]{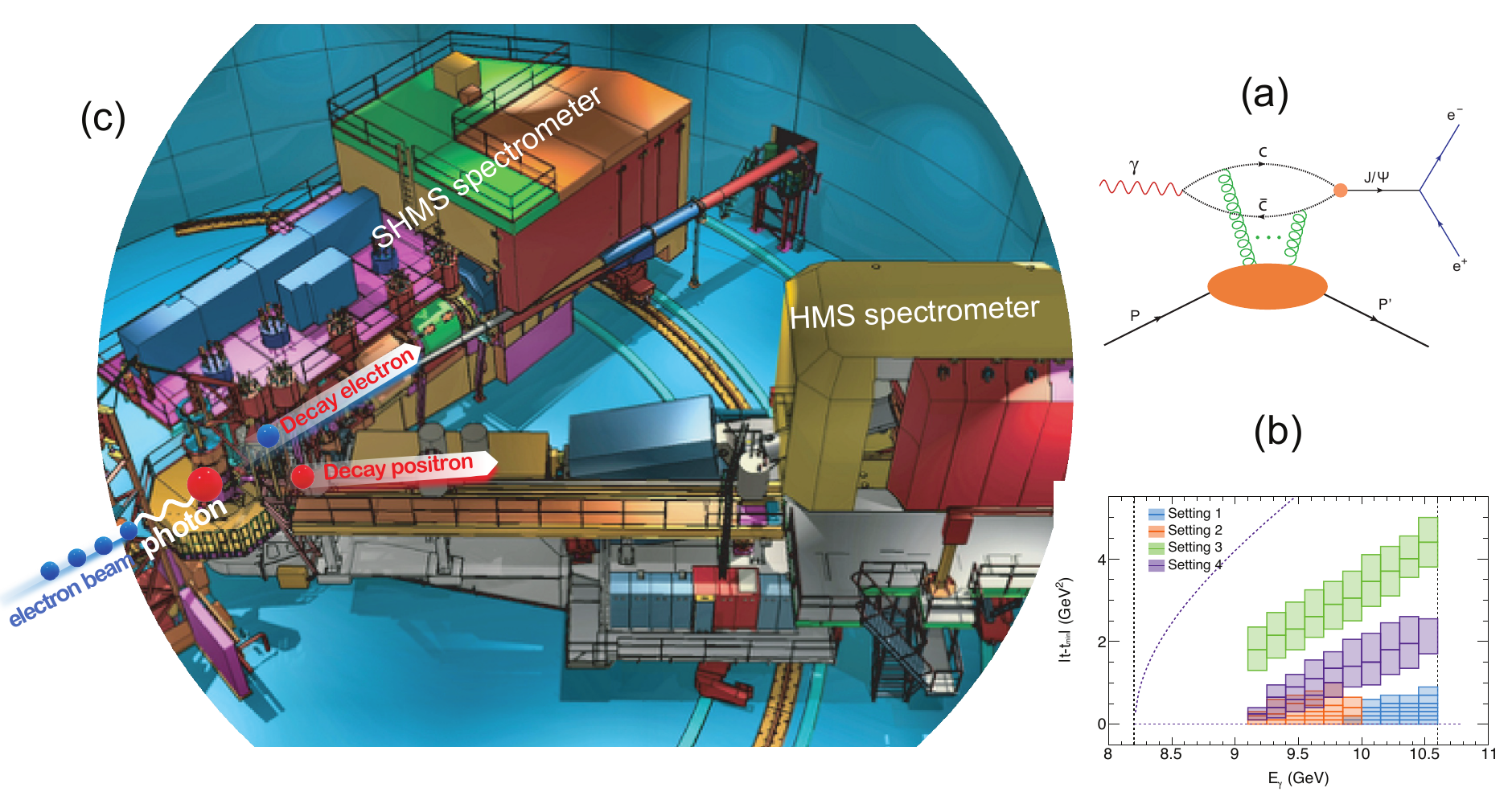}
    \vspace{-0.5cm}
    \caption{{\bf Using a bremsstrahlung photon beam and the decay of the $J/\psi$ to an e+e- pair to measure the production.} \\
    {\bf a)} Standard diagram for exclusive $J/\psi$ photoproduction off a nucleon in the $t$-channel at large invariant mass.  In the threshold region, the exchange involves multiple gluons configurations (denoted by `$...$'). 
    For example, holographic QCD describes the process through tensor 2$^{++}$ graviton-like exchange and scalar 0$^{++}$ exchange. \\
    {\bf b)} Phase space  $\vert t-t_\text{min} \vert$ vs $E_{\gamma}$ , covered in the experiment in four kinematic settings. The dash black vertical lines mark the $J/\psi$ threshold photon energy and the maximum beam energy, 10.6 GeV. The limit values of $t=t_\text{min}$ and $t=t_\text{max}$ are represented by the magenta dash horizontal line, and dash curve respectively. \\
    {\bf c)} Layout of the $J/\psi$-007 experiment in Hall C. A 10.6 GeV incident electron beam strikes a liquid hydrogen target after passing through an 8.5\% Cu radiator. The resulting bremsstrahlung photon beam together with the incident electron beam produces $J/\psi$ particles, which subsequently decay into $e^+e^-$ pairs, detected in coincidence with the HMS and SHMS spectrometers.}
 \label{fig:layout-HC}
 \end{figure*} 
 
In the past 40 years, we have extensively investigated the electric charge and spin of the proton. For example, we learned how the proton charge and magnetization, carried by the electrically charged moving quarks, are distributed and determined the proton electric charge radius through elastic electron scattering~\cite{Xiong:2019umf}. In contrast, the description of the mass distribution of the proton, carried mainly by gluons and their color interactions, is a subject in its infancy: Gluons carry no charge and thus are not amenable to direct study using an electromagnetic probe. 

Recently, it was suggested that measurements of the differential photoproduction of
  $J/\psi$ cross-section in the threshold region as a function of the momentum transfer
  $t$ offer a promising path to access the gluonic GFFs of the
  proton~\cite{Kharzeev:2021qkd,Guo:2021ibg,Hatta:2018ina,Hatta:2018sqd,Hatta:2019lxo,Mamo:2019mka,Mamo:2022eui,Ji:2020bby,Sun:2021gmi}.
  The gluonic GFFs or EMT form factors provide information on the mass, pressure, and
  shear distributions of gluons in the proton~\cite{Pagels:1966zza}. With sufficient data
  at different photon energies, the slopes and magnitudes of the GFFs determined at $t=0$
  give access to the mass radius and potentially to the matrix element of the trace
  anomaly of the energy-momentum tensor. The latter is the ultimate reason for the large
  nucleon mass fraction carried by the gluonic fields. This work reports on a $J/\psi$ photoproduction measurement in the threshold region performed in Hall C at Jefferson Lab.  We present, for the first time, the measured $t$ distributions of the cross sections as a function of the photon energy in the range of $9.1~ {\rm GeV} \leq E_{\gamma} \leq 10.6 ~{\rm GeV}$ and explore the impact of our data on the determination of the GFFs, the proton mass-radius, and the trace anomaly\cite{Ji:2021mtz}. Our focus is the largely unknown gluonic GFFs, the mass and scalar radii of the proton in relation to its charge radius, and the contribution of the trace anomaly to the proton mass~\cite{Ji:2021pys,Lorce:2021xku}.

The near-threshold $J/\psi$ photoproduction experiment, E12-16-007 a.k.a $J/\psi -007$, was carried out at Jefferson Lab between February 8$^{th}$ and March 5$^{th}$, 2019. We measured the exclusive $J/\psi$ photoproduction cross section, depicted in Fig.~\ref{fig:layout-HC}(a), in two dimensions as a function of the Mandelstam variable $t$ and photon beam energy $E_{\gamma}$, covering the phase space shown in Fig.~\ref{fig:layout-HC}(b). We achieved this~\cite{Duran:2021} using a bremsstrahlung photon beam emitted from a 10.6 GeV electron beam while traversing an 8.5\% radiation length Cu radiator positioned about 1~m upstream from the Hall C target chamber. The produced photon beam, together with the electron beam, passed through a target consisting of a 10~cm long cylinder-shaped aluminum can containing liquid hydrogen at 19$^{\circ}$K temperature and 25 psi pressure. We measured the photoproduced $J/\psi$ mesons in the energy range from $E_{\gamma}$= 8.8 GeV up to the bremsstrahlung endpoint energy of $E_{\gamma}$ = 10.6 GeV as shown in Fig.~\ref{fig:layout-HC}(b). The $e^{-}e^{+}$ decay pair of the $J/\psi$ was detected in coincidence using the two high 

\noindent momentum spectrometers of Hall C: the Super High Momentum Spectrometer (SHMS) and High Momentum Spectrometer (HMS) for the electron and the positron, respectively, as shown in Fig.~\ref{fig:layout-HC}(c).

\begin{figure}[H]
\centering
    \includegraphics[width=8.0cm,trim=0 0 0 0,clip]{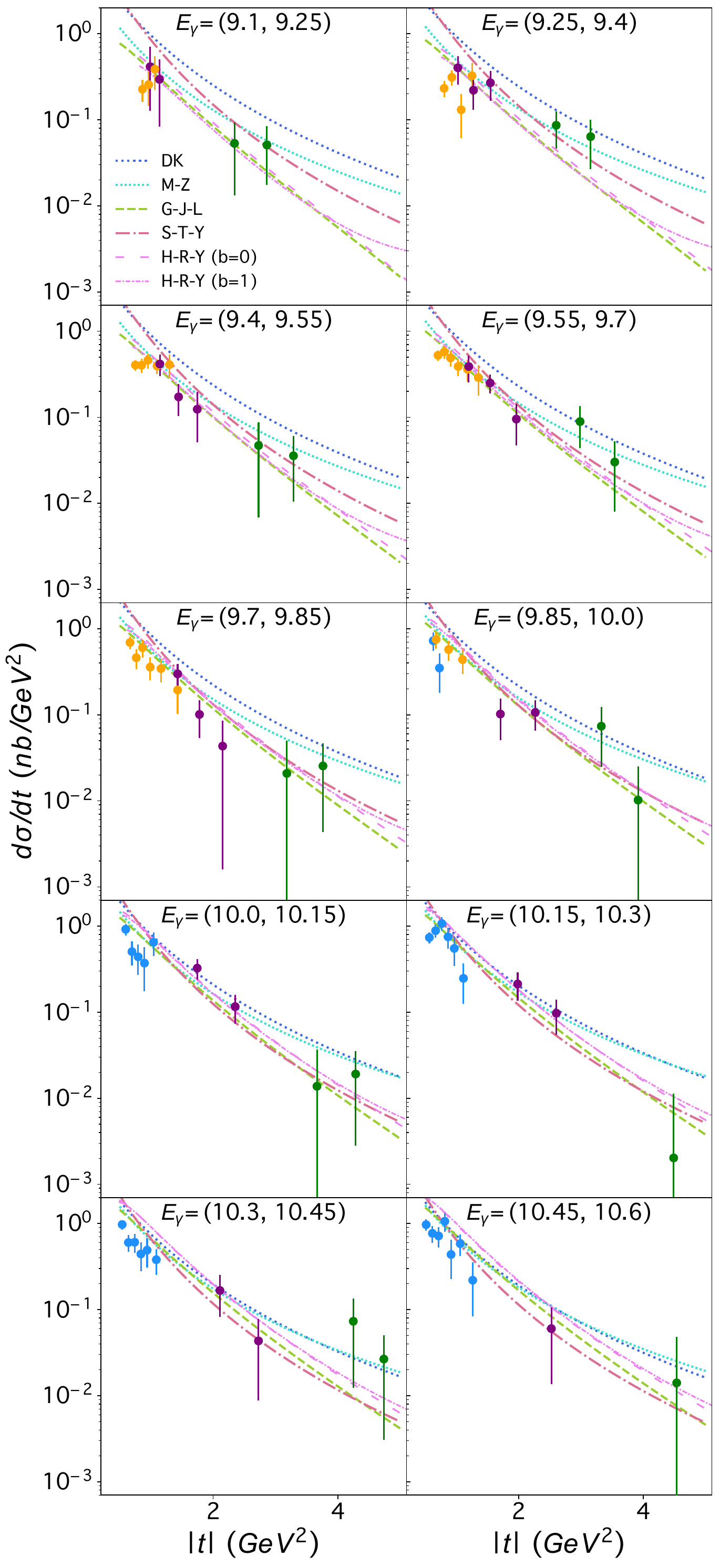} 
    \vspace{-0.1cm}
      \caption{{\bf The differential cross sections versus  $\vert t \vert$.} \\ 
The color of the data points indicates the experimental setting matching the color scheme
  in Fig.~\ref{fig:layout-HC}(b). Each panel shows a different photon energy in GeV with
  0.150~GeV bin size. Every curve is a prediction with fixed parameters determined from
  the GlueX results. The blue dotted line (labeled DK) uses parameters from
  Ref.\citen{Kharzeev:2021qkd}, the cyan dotted line (labeled M-Z) is the holographic QCD
  approach~\cite{Mamo:2019mka}, the green dashed line (labeled G-J-L) is the GPD approach~\cite{Guo:2021ibg}, the red-dash-dotted line is a higher twist approach (labeled S-T-Y)\cite{Sun:2021gmi}, and finally the purple dash (purple dash-dot) labeled H-R-Y is another holographic calculation~\cite{Hatta:2018sqd,Hatta:2018ina,Hatta:2019lxo} with maximal (minimal) trace anomaly contribution to the EMT matrix element.}   
\label{fig:tdistrib}
\end{figure}

In Fig.~\ref{fig:tdistrib}, we show the unfolded two-dimensional differential cross-sections, where each panel corresponds to different central photon energy with a bin width of 150 MeV. The data are compared to the calculations described in references~\cite{Kharzeev:2021qkd,Mamo:2019mka,Guo:2021ibg,Hatta:2018ina,Hatta:2019lxo,Sun:2021gmi} where each model parameters were already published and fixed using the GlueX~\cite{Ali:2019lzf} results at an average photon beam energy of 10.72 GeV with a range from 10 GeV to 11.8 GeV. At photon energies close to the GlueX average photon energy, all models seem to reproduce our data reasonably well but tend to deviate from the data at photon energies below 9.55 GeV, closer to the threshold. One exception is the holographic predictions of Ref.~\citen{Mamo:2019mka} which seems to track the change in the $t$ slope observed in the data.

In order to take advantage of our two-dimensional results,  we expanded our analysis to fit our cross sections using two approaches that explicitly use two GFFs, $A_g(t)$ and $C_g(t)$, in the cross section calculations. Here we assumed that $B_g(t)$'s contribution is small~\cite{Pefkou:2021fni,Mamo:2021krl}. We used both the holographic and generalized parton distribution (GPD) approaches to describe the cross sections to extract the $A_g(t)$ and $C_g(t)$ form factors and deduce {\it one} mass radius and {\it one} scalar radius.

In the holographic QCD calculation of Ref.~\citen{Mamo:2021krl,Mamo:2022eui} (labeled
  M-Z), the dominant exchange is associated with a graviton-like exchange (quantum numbers
  $2^{++}$), however a dilaton-like exchange contribution (quantum numbers $0^{++}$) is
  also included. Both the $A_g(t)$ and $C_g(t)$ form factors are used in the differential
  cross section expression. While these gravitational form factors have a well-defined
  expression~\cite{Mamo:2021krl,Mamo:2022eui} in the holographic calculation, tripole
  approximations inspired by the latest lattice calculations Ref.~\citen{Pefkou:2021fni}
  are used. In our M-Z fitting procedure, the GFFs are parameterized with a total of three
  unknown parameters, $m_T$ for the tripole form of $A_g(t)$, and $C_g(0)$ and $m_S$ for
  the tripole form of $C_g(t)$. The fourth parameter, $A_g(0)$, is related to the momentum
  fraction carried by the proton's gluons, a value well-constrained by the experimental
  data in deep-inelastic scattering. We fixed $A_g(0)$ to the value obtained from the CT18
  global fit~\cite{Hou:2019efy} of the parton distribution functions (PDFs), where $\langle x \rangle_g = 0.414\pm 0.008$. 
Values from other contemporary global fits were also considered and were consistent within one sigma of their uncertainty. Furthermore, it is worth noting that the CT18 value of ${\langle x \rangle}_g$ agrees well with different lattice calculations~\cite{Pefkou:2021fni,Alexandrou:2020sml,Yang:2018nqn}, albeit with a better uncertainty. Finally, the normalization constant in the M-Z approach $\mathcal{N}=7.768$GeV$^{-4}$ was taken from Ref.~\citen{Mamo:2019mka}

In the GPD approach of Ref.~\citen{Guo:2021ibg}, the authors used two GFFs $A_g(t)$ and $C_g(t)$ of a dipole form, fixed the $A_g(0)$ and $m_C$ parameters to lattice~\cite{Shanahan:2018pib}.
Here, as described in the previous paragraph,  we chose tripole forms for both $A_g(t)$
and $C_g(t)$

\noindent while fixing $A_g(0)$ to the value from CT18 PDFs extraction. We then determined $m_A$, $m_C$, and $C_g(0)$ by performing a two-dimensional fit of our data.

\begin{figure}[H]
\centering
\includegraphics[width=8.2cm,trim=0 0 0 0,clip]{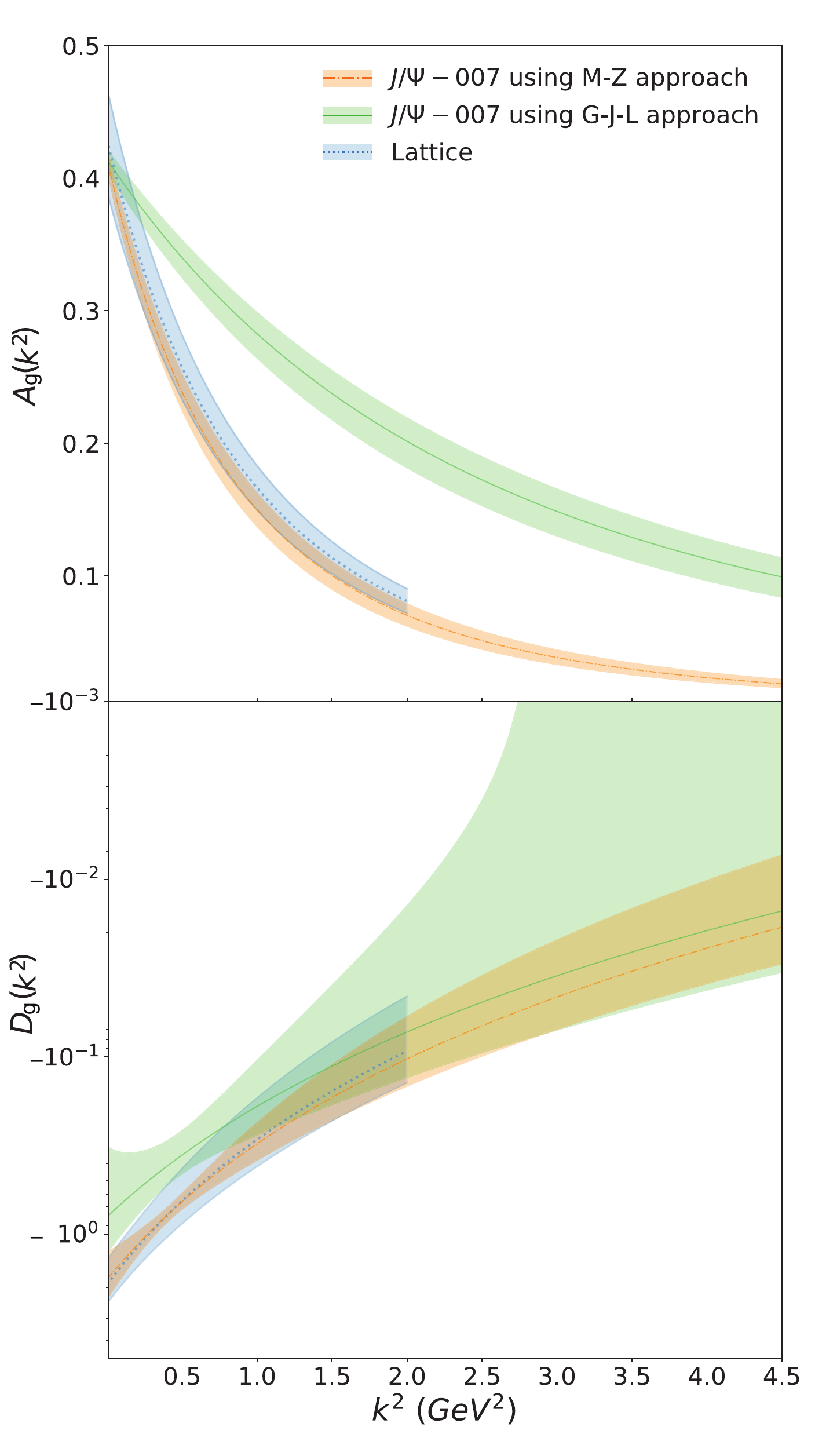}
\caption{ {\bf The gluonic gravitational form factors.}\\
Top panel: The $A_g(k^2)$ form factor extracted from our two-dimensional cross section data in the holographic QCD approach~\cite{Mamo:2021krl} (orange dash-dot curve) and in the GPD approach~\cite{Guo:2021ibg} (green solid curve), compared to the latest lattice calculation~\cite{Pefkou:2021fni} (blue dotted curve). All form factors used the tripole approximation form. The shaded areas show the corresponding uncertainty bands. Bottom panel: The extracted $D_g(k^2)=4C_g(k^2)$ form factor with the same color scheme as the top panel. Note the good agreement between the GFFs from the holographic QCD extraction and the lattice results.}
\label{fig:ACFF}
\end{figure}

In Fig.~\ref{fig:ACFF} both the $A_g(k^2)$ and $C_g(k^2)$ gluonic GFFs
extracted from our two-dimensional cross section data, within the holographic and GPD approaches, are compared to the latest lattice calculation from Ref.~\citen{Pefkou:2021fni}, where $k^2 \equiv \vert t\vert$. For the $A_g(k^2)$ form factor (top panel), the GPD approach results in a form factor that is larger over the entire $t$-range than both the holographic QCD or lattice result. Our data clearly expand the $t$  range up to 4.5 GeV$^2$, doubling the range of $\vert t \vert $ compared to the lattice calculations or extractions from the one-dimensional GlueX results~\cite{Ali:2019lzf}. Furthermore, the two-dimensional nature of our results allows, for the first time, to simultaneously constrain all three unknown parameters, $m_A$, $m_C$, and $C(0)$, from experimental data. The $A_g(t)$ GFF extracted using the holographic QCD approach agrees well with the lattice results, hinting that this approach may provide a path for extracting the gluonic GFFs in this non-perturbative region of $J/\psi$ threshold production.

Finally, after determining the gluonic GFFs in the holographic QCD and GPD approaches, the proton mass and scalar radii can be deduced according to,
\begin{equation}
    \langle r_m^2\rangle_g = 6 \frac{1}{A_g(0)}\frac{dA_g(t)}{dt} {\vert} _{t=0} - 6 \frac{1}{A_g(0)} \frac{C_g(0)}{M^2_N}
\label{radone}
\end{equation}
\begin{equation}
        \langle r_s^2\rangle_g = 6 \frac{1}{A_g(0)}\frac{dA_g(t)}{dt} {\vert} _{t=0} - 18 \frac{1}{A_g(0)} \frac{C_g(0)}{M^2_N}
\label{radtwo}
\end{equation}
where $M_N$ is the nucleon mass. While the above expressions are consistent with the holographic QCD approach, they are approximate in the case of the GPD approach.
Table~\ref{jpsi:fitparams-one} shows the extracted parameters that define the gluonic GFFs for each of the holographic and GPD approaches together with the corresponding mass and scalar radius determined using Eqs.(\ref{radone},\ref{radtwo}). We also report the lattice values for comparison. The radii extracted following the holographic QCD approach, which agree well with the lattice results, suggest a three-fold nucleon structure with a mass radius smaller than the charge radius and a scalar radius extending well beyond the charge radius, about one fermi.

\begin{table*}[t]
\caption{{\bf The gluonic GFF fit parameters and proton mass radius and scalar radius.} \\
They were determined from our data through a two-dimensional fit using the holographic QCD
  and GPD approaches. The corresponding proton mass and scalar radii are also shown
  according to eq.~\ref{radone} and \ref{radtwo}. Note the similar $\chi^2$ per degree of
  freedom in both cases. In all cases we used the tripole-tripole functional form approximation for the GFFs. We compare these results to the latest lattice calculations~\cite{Pefkou:2021fni}.}
\begin{adjustbox}{width=\linewidth}
\begin{tabular}{ccccccc}
\toprule
Theoretical approach & $\chi^2$/n.d.f &$m_A$ (GeV) &  $m_C$ (GeV) & $C_g(0)$ &$\sqrt{\langle r_m^2\rangle}_g$ (fm) & $\sqrt{\langle r_s^2\rangle}_g$ (fm)\\
GFF functional form  &  & & & & \\
\midrule
Holographic QCD  & 0.925 &1.575$\pm$0.059 & 1.12$\pm$0.21 & -0.45$\pm$0.132 & 0.755$\pm$0.035 & 1.069$\pm$0.056  \\
Tripole-tripole  & && & & &\\ 
\midrule
GPD & 0.924 &2.71$\pm$0.19 & 1.28$\pm$ 0.50 & -0.20 $\pm$ 0.11 & 0.472$\pm$0.042 &0.695$\pm$0.071 \\
Tripole-tripole & & & & \\
\midrule
Lattice & & 1.641$\pm$ 0.043 & 1.07$\pm$ 0.12 & -0.483$\pm$ 0.133 & 0.7464$\pm$0.025 &1.073$\pm$0.066 \\
Tripole-tripole & & & & \\
\bottomrule
\end{tabular}
\end{adjustbox}
\label{jpsi:fitparams-one}
\end{table*}

For completeness, we also carried out a one-dimensional analysis of our results to
determine the mass radius according to the vector meson dominance (VMD)-based approach from Ref.~\citen{Kharzeev:2021qkd} (labeled DK) for each photon energy separately. 
Additionally, following the VMD approach described in
refs.~\cite{Wang:2019mza,Kharzeev:1995ij,Kharzeev:1998bz,Kharzeev:2021qkd}, we extracted
the anomalous mass contribution to the proton mass $M_a/M$ comparing an exponential GFF
(as in Ref.~\citen{Wang:2019mza}) to a dipole form GFF (as in
Ref.~\citen{Kharzeev:2021qkd}). This analysis is discussed in the Methods section.

We have measured the exclusive $J/\psi$ photoproduction differential cross sections near threshold as a function of photon energy $E_{\gamma}$ from 9.1 GeV to 10.6 GeV, and four-momentum transfer $t$ from $t_{min}$ up to 4.5 GeV$^2$.  We fit our two-dimensional data  using the cross section expressions from the holographic QCD approach~\cite{Mamo:2021krl} and the GPD

\noindent approach~\cite{Guo:2021ibg} to fully determine, for the first time, the gluonic GFFs $A_g(t)$ and $C_g(t)$ from purely experimental data, up to $\vert t \vert=4.5 \text{GeV}^2$. We found $Dg(t)=4C_g(t)$ to be negative over the measured $t$-range. We compared our extracted  results to lattice QCD~\cite{Pefkou:2021fni} and find a very good agreement when  we followed the holographic QCD approach. Using our results,  we derived the proton mass radius and scalar radius in each approach, finding the proton mass radius to be smaller than
its charge radius. Furthermore, the holographic QCD 
extraction yields a scalar radius of one fermi, substantially larger than the charge radius. This compels us to see the proton's structure as consisting of three distinct regions. An inner core, dominated by the tensor gluonic field structure, provides most of the proton's mass. The charge radius, determined by the relativistic motion of the quarks, extends beyond this inner core. The entirety of the proton is enveloped in a confining scalar gluon density, extending well beyond the charge radius.

\bibliography{sn-bibliography}

\section*{Methods}
An in-depth overview of the experimental methodology for this work can be found in Ref.~\citen{Duran:2021}.

\subsection*{Design}

We used the Hall C spectrometers, shown in Fig.~\ref{fig:layout-HC}(c), in four complementary configurations (see 
Table~\ref{jpsi:settings}), 
to measure and identify charged particles. The detector packages of each spectrometer comprise two sets of identical drift chambers for tracking purposes, two pairs of XY-plane hodoscopes, a threshold Cherenkov counter, and an electromagnetic calorimeter consisting of a pre-shower and a shower unit. The hodoscopes are used as the trigger in each spectrometer. The single-particle trigger rate for each spectrometer varied from 10 to 200~KHz depending on the central angle and momentum settings of the spectrometers. The coincidence trigger is formed by the overlap of the SHMS and HMS single-particle triggers within a narrow time window. 
The time resolution of the hodoscopes was sufficient to resolve the 2 ns beam bunch structure. 
Since the single-particle trigger did not include particle identification, we measured a combination of electrons (positrons), muons, and charged pions in both spectrometers. The coincidence trigger collected the $e^{+}e^{-}$ decay pairs from the photoproduced $J/\psi$s and the combinatoric background from $e^-h^+$ and $h^-h^+$ coincidence events. In turn, this enabled the simultaneous measurement of the background contributions to the $J/\psi$ mass peak. Furthermore, the data include $\mu^{+}\mu^{-}$ decay pairs data from photoproduced $J/\psi$s; the results from this measurement will be described in a separate publication.

\subsection*{Reconstruction}
 
We selected electrons with the SHMS using the electromagnetic calorimeter and positrons with the HMS using the Cherenkov counter and electromagnetic calorimeter. 
We reconstructed the particle momenta and positions at the interaction vertex from the positions and angles determined at the spectrometer focal plane, measured with two sets of drift chambers, using the optical properties of the spectrometers.
We corrected for tracking inefficiencies in each spectrometer as a function of the spectrometer's trigger rate. To account for the computer and electronics livetime, we used the Hall C electronics deadtime monitoring system (EDTM), where a fixed frequency pulse is inserted in the HMS and SHMS trigger logic to mimic real pre-triggers signals. This resulted in a rate-dependent EDTM correction. 
We also considered and corrected the effect of target density loss due to temperature fluctuations in the liquid hydrogen target.
Finally, due to the low positron rate in the HMS compared to the positive hadron rates, it was necessary to use a different reaction to study the efficiency of the coincidence measurement between both spectrometers. This was possible using $e^-\pi^+$ background events. We instituted a rate-dependent correction based on the measured inefficiencies for these events.

We determined both spectrometer acceptances using a forward single-arm Monte-Carlo
simulation (SAMC~\cite{Duran:2021}), accounting for the target geometry, spectrometer
optics, and detector resolutions (including the impact of internal and external radiative
effects on the electrons and positrons that occure before and after the interaction
point). The simulation used the empirical fit~\cite{Bosted:2007xd} to world
electron-nucleus inclusive scattering cross section data (using the fit designated as
"F1F2-21"). In the kinematic region of our inclusive electron-proton scattering calibration data, the
empirical fit has a systematic uncertainty of 3\%. The HMS has been used for many
experiments in Hall C over the past 30 years, hence we found the acceptance to be
well-described by the existing Monte-Carlo simulation of the detector. The SHMS was installed in Hall C in 2017 and understanding of its acceptance is ongoing. We found that in the central region of momentum and target position, the data and simulation agreed well, while we observed larger disagreements at the edges, with smooth dependence on momentum and target position. We corrected this with momentum and position-dependent acceptance factor.
Finally, we calibrated the central momentum and angles of the spectrometers by measuring elastic scattering on the hydrogen target.

We were able to cleanly measure the $J/\psi$ invariant mass spectra by combining the event selection criteria described above with a strict timing requirement based on the trigger time and the travel time of the decay leptons from the target to the detectors. Due to the relatively large angles of the spectrometers and the good experimental resolution, the invariant mass peaks are very prominent against a low background.
Contamination of the invariant mass spectra from $e^-\pi^+$ is the dominant physics background.  We were able to precisely determine the background shape through our direct measurement of these events, where we fixed the background normalization to the sidebands of the invariant lepton pair mass spectrum.
After background subtraction, we obtained the raw $J/\psi$ cross section by integrating the remaining spectrum over a fixed window around the $J/\psi$ invariant mass, as a function of the photon energy $E_\gamma$ and Mandelstam variable $t$.

The observed raw $J/\psi$ cross section consists of events from bremsstrahlung photons and
quasi-real photons from the electron beam, interacting with either the liquid hydrogen
target or with the aluminum target entrance and exit windows.
We removed the contributions from the target windows based on measurements with a two-aluminum-foil target.
To account for contributions from quasi-real events, we determined the fraction of electroproduction events to photoproduction events using the lAger Monte Carlo generator\cite{Gryniuk:2020mlh} and a full simulation of the experimental setup.
We then unfold for radiative effects on the $J/\psi$ decay, limited detector resolution and acceptance, the bremsstrahlung photon flux, and the $J/\psi$ branching ratio through an iterative unfolding approach~\cite{DAgostini:1994fjx} using two Monte-Carlo samples.
The resulting Born-level photoproduction cross section for each experimental setting is differential in $t$ as a function of $E_\gamma$.

\subsection*{Backgrounds}

Since we did not detect the recoil proton, the degree of exclusivity of the reaction was
aided by its near-threshold kinematics and the narrow acceptance of the spectrometers.
Background events, where an extra pion is produced with the $J/\psi$ in the range of
available photon energies, would fall outside the kinematic settings for the elastic
$J/\psi$ production and would reconstruct below $t_\text{min}$. However, this event
signature was not observed in our data sample.  
Another background we investigated in our kinematics settings is the Bethe-Heitler
process. This potential background contribution was evaluated according to Ref.~\citen{Pauk:2015oaa} and found to be small ($< 1$\% ) due to the combined spectrometers very limited acceptance and large spectrometer angles. This is consistent with the lack of an observed shoulder in the $e^{+}e^{-}$ invariant mass spectrum.  

\subsection*{Systematic Uncertainties}

The systematic uncertainties on our two-dimensional cross section consist of two parts: an
overall 4\% scale uncertainty and an additional point-to-point uncertainty. The
point-to-point uncertainties are in all cases substantially smaller than the statistical uncertainties. Finally, we investigated bin-to-bin correlations due to the unfolding procedure and found the effect to be negligible. We evaluated our total uncertainty by adding quadratically the statistical and systematic errors and used it in all of our fitting procedures.

The scale uncertainty is mainly driven by the 3\% uncertainty on the F1F2-21 deep
inelastic scattering model we used to calibrate our detector acceptance. Other major contributions to the scale uncertainty are due the correction for residual rate dependence (1.2\%), to the luminosity measurement (1\%), residual correction to the SHMS acceptance based on the vertex position (1\%), target window subtraction (1\%), subtraction of electroproduction events determined by performing measurements without the Cu radiator (1\%), radiator thickness (1\%), detector livetime correction (0.45\%), tracking efficiency correction (0.44\%), target boiling correction (0.44\%). The systematic uncertainty due to particle identification inefficiencies was also considered and found to be negligible.

We considered the following sources to estimate the point-to-point systematic uncertainties: to estimate the uncertainties due to the background subtraction on the invariant mass, the radiative corrections, and the material effects we conducted the entire cross section analysis with varying integration windows on the invariant mass spectrum, varying between very narrow (cutting out the tail on the $J/\psi$ invariant mass peak), and very wide. The maximum absolute difference between the different variant analyses was taken as the contribution to the systematic uncertainty. To estimate the dependence of the unfolding procedure on the Monte-Carlo model, we repeated the approach with two very different models: a two-gluon model fit to the world data excluding GlueX and assuming $s$-channel helicity conservation, and a different model from the JPAC group~\cite{Mathieu:2018xyc} which includes the recent GlueX results and uses a full description of the spin-density matrix elements to describe the $J/\psi$ polarization. The difference between both models was 

\begin{table}[H]
\caption{{\bf Spectrometers Settings.}\\
The polarity, momentum and angle settings for the used for the $J/\psi$-007 experimental measurements.}
\begin{adjustbox}{width=7cm}
\begin{tabular}{ |c|c|c|} 
\hline
\textbf{Settings} & \textbf{HMS~(+)} &  \textbf{SHMS~(-)}  \\
\hline
Setting 1 & 4.95 GeV, 19.1$^{\circ}$ & 4.835 GeV, 17.0$^{\circ}$  \\
\hline
Setting 2 & 4.6 GeV, 19.9$^{\circ}$ & 4.3 GeV, 20.1$^{\circ}$  \\
\hline
Setting 3 & 4.08 GeV, 16.4$^{\circ}$ & 3.5 GeV, 30.0$^{\circ}$ \\
\hline
Setting 4 & 4.4 GeV, 16.5$^{\circ}$ & 4.4 GeV, 24.5$^{\circ}$  \\
\hline
\end{tabular}
\end{adjustbox}
\label{jpsi:settings}
\end{table}
\begin{figure}[H]
\centering
    \includegraphics[width=8.0cm,trim=0 0 0 42,clip]{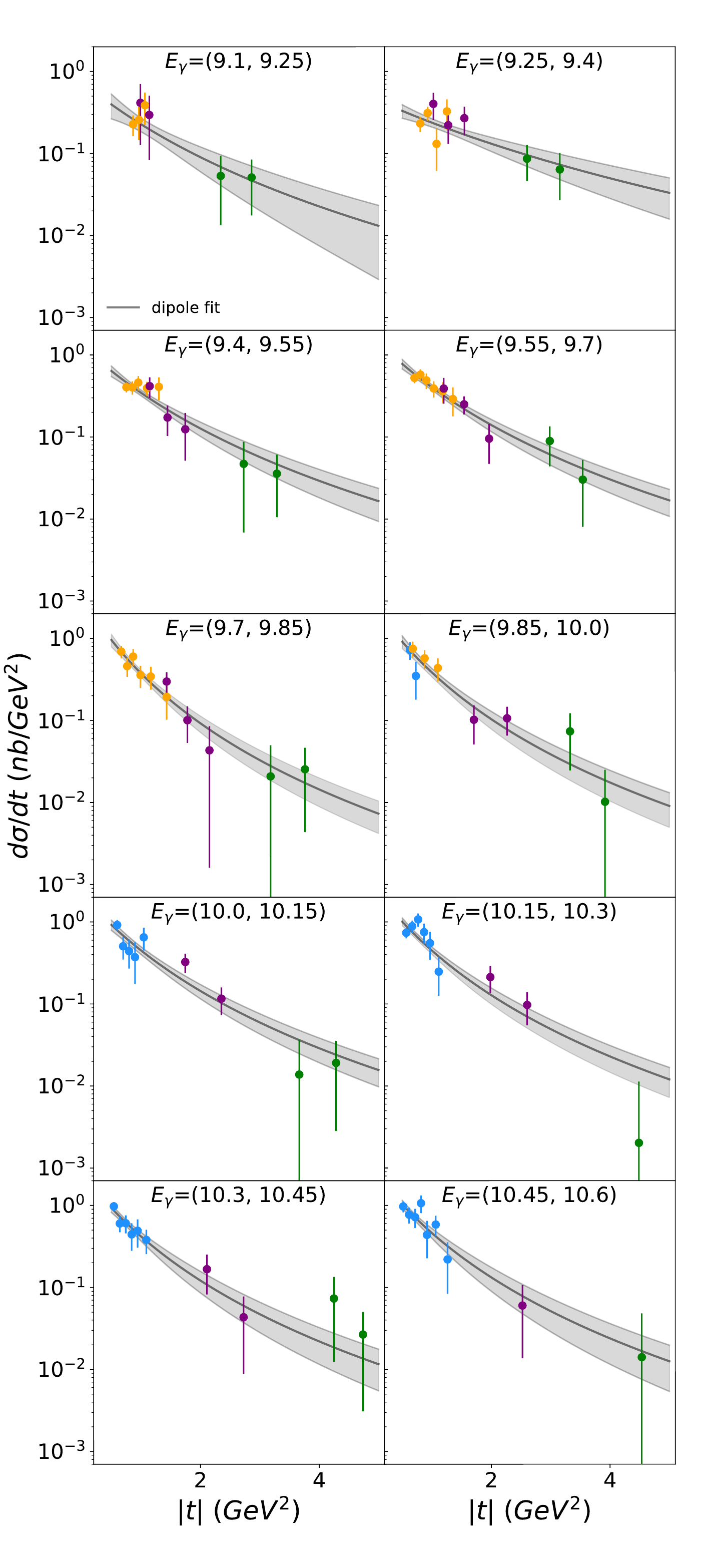} 
    \vspace{-0.6cm}
      \caption{{\bf Fit to differential cross sections versus  $\vert t \vert$.}\\
The differential cross section data are identical to Fig.\ref{fig:tdistrib} and the black solid curve is a dipole fit to the data according to Ref.~\citen{Kharzeev:2021qkd} while the grey band shows its uncertainty. The parameters are listed in Table~\ref{jpsi:fitparams-two}.}   
\label{fig:dfit-tdist}
\end{figure}

\noindent taken as systematic uncertainty. To obtain the total systematic uncertainty we added both contributions in quadrature.

\subsection*{Kinematic settings}

In Table~\ref{jpsi:settings} we summarize each kinematic setting.  We list the spectrometers' momentum and angle settings. Their corresponding coverage in $t$ and $E_{\gamma}$ is shown in Fig.~\ref{fig:layout-HC} (b). 

\subsection*{Interpretation in the DK approach}

In the case of DK approach, we extracted the mass radius first. To this end, we fitted our $t$ dependent differential cross sections and extracted the radius at each photon energy following Ref.~\citen{Kharzeev:2021qkd} prescriptions 

\begin{equation}
\frac{d\sigma}{dt} =\frac{1}{64 \pi s} \frac{1}{\vert p_{\gamma cm} \vert ^2}(Q_ec_2)^2\left(\frac{16\pi^2M}{b}\right)^2 G(t)^2,
\end{equation}

\noindent where $s$ is the square of the invariant mass of the photon-nucleon system, $M$ is the nucleon mass, $p_{\gamma cm}$ is the center of mass photon momentum $Q_e$ the electric charge of the charm quark ($Q_e= 2e/3$), $c_2$ is a short distance coefficient, determined from the fit and $b = 11-2n_l/3 - 2n_h/3 = 9$ with $n_l$ and $n_h$ the number of light and heavy quarks respectively. Here one effective scalar GFF of a dipole form, $G(t)=M(1 - t/m_s^{2})^{-2}$ is used with $G(0) = M$ in the rest frame of the particle. It encompasses a combination of the three GFFs, $A_g(t)$, $B_g(t)$ and $C_g(t)$ of the standard decomposition of the energy-momentum tensor in a nucleon state. The proton mass radius is then given according to Ref.~\citen{Kharzeev:2021qkd} by,
\begin{equation}    
\langle r_m^2 \rangle =\frac{6}{M}\frac{dG}{dt} \vert_{t=0}=\frac{12}{m_s^2}
\label{dradius}
\end{equation}

\begin{figure}[H]
\centering
\includegraphics[width=8cm,trim=0 0.5cm 0 2.0cm,clip]{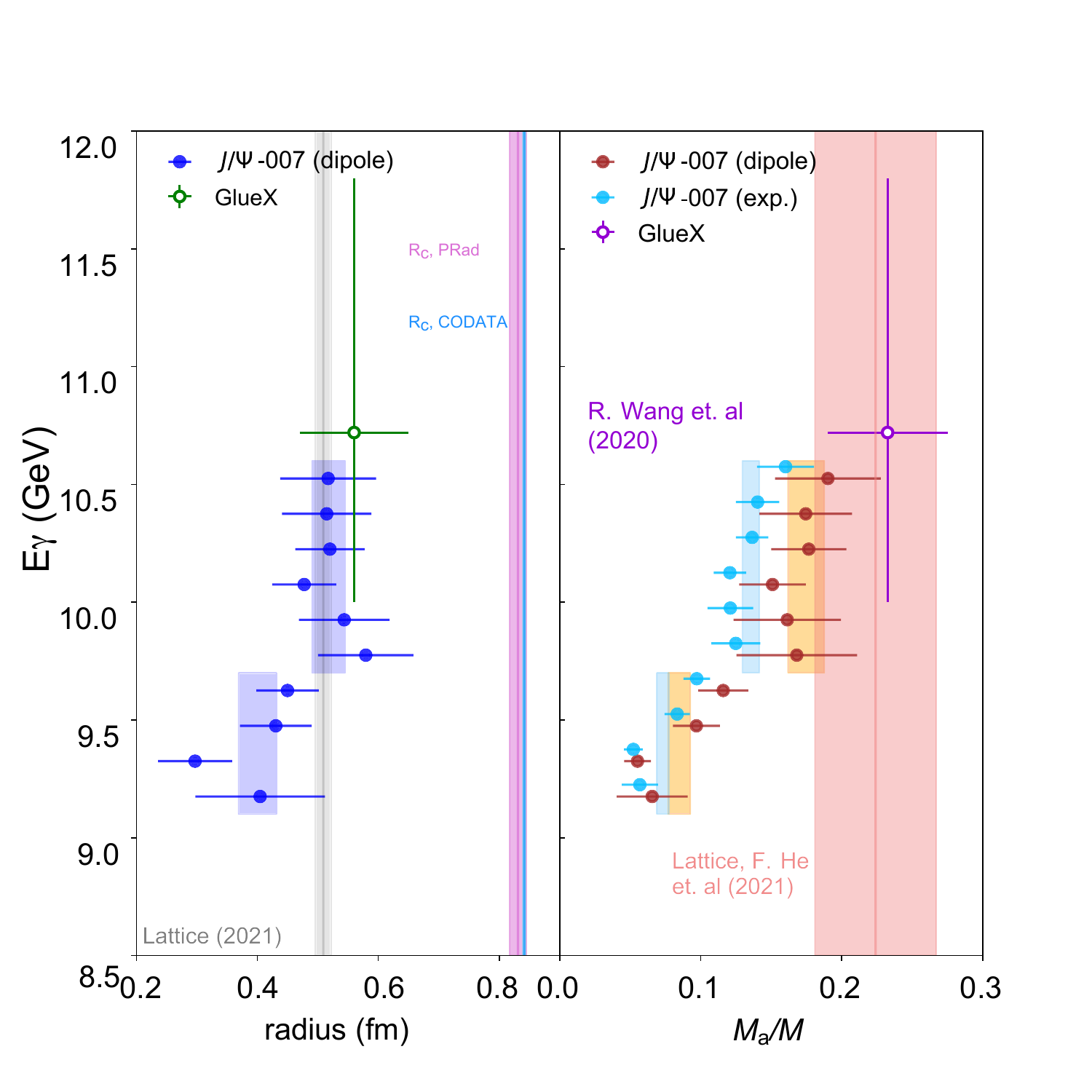}
     \caption{ Left panel:The extracted radius as a function of the photon energy 
according to Ref.~\citen{Kharzeev:2021qkd} together with the GlueX result. Both our and
  the Gluex extractions used a dipole fit of the form factor. The charge radius from
  CODATA and the latest electron scattering~\cite{Xiong:2019umf} (labeled PRAD) are
  plotted as lines with error bands. The lattice result\cite{Pefkou:2021fni} is plotted as a grey line with grey error band. Right panel: The extracted $M_a/M$ according to Ji's mass decomposition~\cite{Ji:1994av} following ~\cite{Wang:2019mza} along with a recent direct lattice calculation of the same quantity~\cite{He:2021bof}.}
\label{fig:radanomdk}
\end{figure}

The experimental data from our measurement, together with a dipole fit (solid black 
line) according to Ref.~\citen{Kharzeev:2021qkd} and its
corresponding uncertainty band (gray shaded area) are presented in 
Fig.~\ref{fig:dfit-tdist}.
The extracted mass radii at different photon energies, according to this prescription, are shown in 
Fig.\ref{fig:radanomdk} (left). 
At higher energies, an energy-independent region of radii consistent with GlueX emerges, however, we see a decrease in the size of the extracted radius as we get closer to the threshold, below 9.7 GeV. The radius determined in the energy  independent region averages to $\sqrt{ \langle r^2_m \rangle} = 0.52 \pm 0.03 {\rm fm}$. This result compares favorably with the lattice QCD determination of the radius using solely the $A_g(t)$ gluonic GFF of ~\cite{Shanahan:2018pib,Pefkou:2021fni} and equation~(\ref{dradius}). We observe that in~\cite{Kharzeev:2021qkd} approach the mass radius is clearly smaller than the charge radius of the proton.

In Table~\ref{jpsi:fitparams-two} we list the fit parameters and the corresponding uncertainties in the approach of Ref.~\citen{Kharzeev:2021qkd}. From those parameters, we extract the radius according to~\cite{Kharzeev:2021qkd} and the anomalous energy contribution~\cite{Ji:1994av} to the proton mass according to~\cite{Wang:2019mza}.

Moreover, following the same procedure described in Ref.~\citen{Wang:2019mza}, based also on~\citen{Kharzeev:1995ij,Kharzeev:1998bz,Kharzeev:2021qkd} assuming a dilaton ($0^{++}$) scalar exchange between the $J/\psi$ and the nucleon, we inferred the quantum anomalous energy discussed in Ref.~\citen{Ji:2021mtz}. The results are shown in the right panel of 
Fig.~\ref{fig:radanomdk}. 
We extracted $M_a/M$ using both an exponential GFF as in Ref.~\citen{Wang:2019mza} as well as a dipole form GFF as in Ref.~\citen{Kharzeev:2021qkd}. We used the

\begin{table*}[htb!]
\caption{{\bf Fit parameters, mass radius, and trace anomaly.}\\
The table reports the fit parameters of the dipole fit, the mass radii according to DK~\cite{Kharzeev:2021qkd} and the trace anomaly according to~\cite{Wang:2019mza} at each photon energy.}
\begin{adjustbox}{width=\textwidth}
\begin{tabular}{ccccccc}
\toprule
$E_{\gamma}$(GeV)&$m_A$ (GeV) & $c_2$ (nb/GeV$^2$)$^{\frac{1}{2}}$ & $\chi^2/n.d.f$ & Mass radius (fm) & $M_a/M$(dipole) &  $M_a/M$(exp.)\\
\midrule
9.10-9.25 & 1.69$\pm$0.45 &  34.10$\pm$10.60 & 0.45 & 0.404 $\pm$ 0.107 &  0.066 $\pm$ 0.025 &  0.057 $\pm$ 0.013 \\
9.25-9.40 & 2.30$\pm$0.47 &   27.37$\pm$4.30 & 0.96 & 0.297 $\pm$ 0.061 &  0.055 $\pm$ 0.009 &  0.052 $\pm$ 0.007 \\
9.40-9.55 & 1.59$\pm$0.22 &   46.34$\pm$7.35 & 0.89 & 0.430 $\pm$ 0.059 &  0.097 $\pm$ 0.017 &  0.083 $\pm$ 0.009 \\
9.55-9.70 & 1.52$\pm$0.17 &   53.57$\pm$7.63  & 0.39 & 0.450 $\pm$ 0.052 &  0.116 $\pm$ 0.018 &  0.097 $\pm$ 0.009 \\
9.70-9.85 & 1.18$\pm$0.16 &  75.39$\pm$16.88 & 0.49 & 0.579 $\pm$ 0.079 &  0.168 $\pm$ 0.043 &  0.125 $\pm$ 0.017 \\
9.85-10.00 & 1.26$\pm$0.17 &    70.06$\pm$14.65 & 0.85 & 0.544 $\pm$ 0.075 &  0.161 $\pm$ 0.038 &  0.121 $\pm$ 0.016 \\
10.00-10.15 & 1.43$\pm$0.16 &   63.75$\pm$9.24 & 1.20 & 0.477 $\pm$ 0.053 &  0.151 $\pm$ 0.024 &  0.121 $\pm$ 0.011 \\
10.15-10.30 & 1.31$\pm$0.14 &   72.74$\pm$10.20 & 2.01 & 0.520 $\pm$ 0.057 &  0.177 $\pm$ 0.027 &  0.136 $\pm$ 0.011 \\
10.30-10.45 & 1.33$\pm$0.19 &   70.29$\pm$12.07 & 0.48 & 0.515 $\pm$ 0.074 &  0.174 $\pm$ 0.033 &  0.140 $\pm$ 0.015 \\
10.45-10.6 & 1.32$\pm$0.20 &  75.32$\pm$13.59  & 0.74 & 0.517 $\pm$ 0.079 &  0.190 $\pm$ 0.038 &  0.160 $\pm$ 0.020 \\
\bottomrule
\end{tabular}
\end{adjustbox}
\label{jpsi:fitparams-two}
\end{table*}

\noindent energy independent region to determine an average value and find $M_a/ M = 0.175\pm 0.013$. Of course, this quantity should be energy independent but clearly, it is not as we consider photon energies smaller than 9.7 GeV and thus raises the question about the validity of its interpretation. Nevertheless, it is interesting that the dipole form factor extraction is closer to the first direct lattice calculation~\cite{He:2021bof} of the same quantity and offers a glimpse on the origin of the proton mass. 

\section*{Data availability}
The raw data from the experiment are archived in Jefferson Laboratory mass storage silo
and at Argonne National Laboratory. The analyzed data are archived at Argonne National
Laboratory. The data are available in the Supplemental Material and a CSV file of the
cross section data is available in the Supplementary Data.

\section*{Acknowledgements}
This work was supported in part by the US Department of Energy Office of Science, Office of Nuclear Physics under contracts numbers DE-AC02-06CH11357 and  DE-FG02-94ER40844, including contract number DE-AC05-06OR23177, under which Jefferson Science Associates, LLC operates the Thomas Jefferson National Accelerator Facility.

\section*{Competing interests} 
The authors declare no competing interests. 

\section*{Correspondence and request of materials}
Correspondence and request of materials should be addressed to Z.-E.M.

\section*{Author contributions}
  S. Joosten, M.K. J., Z.-E. M., M. P., and E. C. are co-spokespersons of the experiment. The data analysis was carried out by  B. D, S. J., M. K. J., S. P., C. P., and Z.-E. M. All authors reviewed the manuscript. The entire $J/\psi$-007 collaboration participated in the data collection and in the online analysis of the experiment.

\end{multicols}

\end{document}